\documentclass[aps,prd,twocolumn,floatfix,nofootinbib,superscriptaddress,
showpacs,amssymb]{revtex4}

\usepackage{bm}
\usepackage{graphicx}

\def\bea{\begin{eqnarray}}
\def\eea{\end{eqnarray}}
\def\beq{\begin{equation}}
\def\eeq{\end{equation}}
\def\p{\partial}
\def\vp{\varphi}

\newcommand{\w}[1]{\bm{#1}}

\begin{document}

\title{The final phase of inspiral of strange quark star binaries}

\newcommand*{\MEU}{Laboratoire de l'Univers et de ses Th\'eories, UMR 8102
  du C.N.R.S., Observatoire de Paris, F-92195 Meudon Cedex, France}
\newcommand*{\COR}{Center for Radiophysics and Space 
Research, Cornell University, Ithaca, New York, 14853, USA} 
\newcommand*{\UNI}{Institute of Astronomy, University of Zielona G\'ora, Lubuska 2, 65-265, Zielona G\'ora,  Poland} 
\newcommand*{\ALI}{Departament de Fisica Aplicada, Universitat d'Alacant, Apartat de correus 99, 03080 Alacant, Spain}

\author{Dorota Gondek-Rosi\'nska}\email{Dorota.Gondek@obspm.fr}\affiliation{\UNI,\MEU, \ALI}
\author{Fran\c cois Limousin}\email{limousin@astro.cornell.edu}\affiliation{\COR, \MEU}

\date{30 January 2008}

\begin{abstract}
We present calculations of the final phase of inspiral of irrotational
strange star binaries. Two types of equation of state at zero
temperature are used - the MIT bag model and the Dey et al. 1998 model
of strange quark matter. We study the precoalescence stage within the
Isenberg-Wilson-Mathews approximation of General Relativity using a
multidomain spectral method. The gravitational-radiation driven
evolution of the binary system is approximated by a sequence of
quasi-equilibrium configurations at a fixed baryon number and with
decreasing separation. We find that the innermost stable circular
orbit (ISCO) is determined always by an orbital instability for
binaries consisting of two stars built predominantly of strange quark
matter independently on the total mass of a binary system and
compactness parameter of each star.  In contrast, for neutron stars
described by baryonic equation of state without exotic phases the ISCO
is given by the mass-shedding limit.  The gravitational wave frequency
at the ISCO, which marks the end of the inspiral phase, is always
higher than 1.1kHz for equal masses irrotational strange quark stars
with the total mass-energy of a binary system greater than $2
M_\odot$. We find that the dependence of the frequency of
gravitational waves at the ISCO on the compactness parameter for the
equal mass binaries can be described by the same simple analytical
formulae for broad ranges of masses independently on a strange star
model. Detailed comparisons with binary neutrons star models, as well
as with the third order Post-Newtonian point-mass binaries are given.
The difference in the phase, for two $1.35 M_\odot$ strange stars,
between our numerical results and 3PN is $\sim 40 \%$ for the last two
orbits of inspiral.

\end{abstract}


\maketitle

\section{Introduction}
Coalescing compact object binaries are the strongest and hence the most
promising sources of gravitational waves (GW) for LIGO, VIRGO and other
interferometric detectors \cite{Burga03, Kalogera04, Belczynski02}. Among
these, binary neutron stars have been a subject of extreme interest since the
GW signal of terminal phases of evolution of such binary system could yield
important information about the equation of state (EOS) at nuclear densities
(e.g \cite{Lai1996,Faber02,TanigG03,Gondek07,LimouGG05,Oechslin04,Bejger05,ShibaTU05}). One
can impose constraints on the EOS of neutron stars using a simple method
based on the properties of quasiequilibrium binary sequences \cite{Faber02,
Bejger05, LimouGG05}. The individual masses of the two neutron
stars in a binary system can be determined taking into account the frequency
evolution of the GW signal of the inspiral phase and high-order PN effects on
the phase evolution of the signal \cite{CutleF94}. In addition 
 the compactness parameter $M/R$, hence $R$, of neutron
stars (where $M$ is gravitational mass and $R$ stellar radius of an isolated
neutron star) can be found  based on the observed
deviation of the gravitational energy spectrum of a quasiequilibrium sequence
from point-mass behavior at the end of inspiral \cite{Faber02}.

Several groups have studied the last orbits of inspiraling binary neutron
stars in the quasi-equilibrium approximation, and in the framework of
Isenberg-Wilson-Mathews (IWM) approximation of general relativity (see
\cite{BaumgS03} for a review). The quasi-equilibrium assumption
approximates the evolution of the system by a sequence of exactly circular
orbits (as the time evolution of the orbit is much larger than the orbital
period). The IWM approximation amounts to using a conformally flat spatial
metric, which reduces the problem to solving only five of the ten Einstein
equations.   The equilibrium configurations have been
calculated for irrotational binaries since the viscosity of neutron star
matter is far too low to ensure synchronization during the late stage of the
inspiral \cite{BildsC92,Kocha92}.

In order to construct accurate templates of expected GW signal from neutron
stars binaries one has to take into account realistic description of nuclear
matter and astrophysically relevant masses of neutron stars in a binary system.

Almost all relativistic studies of the final phase of the inspiral of close
binary neutron stars systems employ a simplified EOS of dense matter, namely a
polytropic EOS \cite{BonazGM99,MarroMW99,UryuE00,UryuSE00,GourGTMB01,Faber02,
TanigG02b,TanigG03,FaberGR04,MarronDSB04}. There are only three exceptions:
(i) Oechslin et al. have used a pure nuclear matter EOS, based on a
relativistic mean field model and a `hybrid' EOS with a phase transition to
quark matter at high density \cite{Oechslin04}; (ii) Bejger et al. have
computed quasi-equilibrium sequences based on three nuclear matter EOS
\cite{Bejger05} iii) Limousin et al. have studied the properties of binary
strange quark stars described by the simplified MIT bag model (with massless
and not interacting quarks) of strange quark matter \cite{LimouGG05}. In these
three papers the authors have considered only binary systems consisting of two
identical stars. The assumption of almost equal masses of neutron stars in
a binary system was based on the current set of well-measured neutron stars
masses in relativistic binary radio pulsars. However, one has to note that the
conclusions based on analysis of properties of radio binary pulsars suffer
from small number statistics and from several selection effects \cite{GondeBB05,BulikGB04}.

In this article we calculate the final phase of
inspiral of irrotational strange quark star binaries using the MIT bag model
and the Dey et al. (1998)\cite{Dey98} model of strange quark matter. We study the impact of
the equation of state and total energy-mass on the last orbits of binary
strange quark stars.  We compare the evolution of strange star binaries with
neutron star binaries in order to find any characteristic features in the
gravitational signal that will help to distinguish between strange stars and
neutron stars.

The paper is organized in the following way: Sec. II is devoted to the
description of the EOSs used to describe strange stars. Sec. III is a brief
summary of the assumptions upon which this work is based and a short 
description of the basic equations for quasi-equilibrium configurations. 
In Sec. IV we comment on the properties of evolutionary sequences and define 
the notion of innermost stable circular orbit.
In Sec. V we present the numerical results for irrotational strange stars
binaries of $M=1.35 M_\odot$ and compare their evolution with that of 
neutron stars. Then in Sec. VI we show the results obtained for binary strange 
stars with different total mass and compare them with neutron star binaries.
Section VII contains the final discussion.

Throughout the paper, we use geometrized units, for which $G=c=1$, where $G$
and $c$ denote the gravitational constant and speed of light respectively.

\section{Equations of state and stellar models}

\begin{table}[h]
\begin{center}
\begin{tabular}{|c|c|c|c|c|}
\hline
\multicolumn{1}{|c}{EOS} & \multicolumn{1}{|c|}{$M [M_{\odot}]$} &
\multicolumn{1}{|c|}{$M_{\rm B} [M_{\odot}]$}&
\multicolumn{1}{|c|}{$R~[{\rm km}]$}&
\multicolumn{1}{c|}{$M/R$} \\
\hline
\hline
                       &      &        &              &\\
SQSB40                 & 0.5  & 0.5611 & $\ $9.001$\ $& $\ $0.0820$\ $  \\
                       & 1.35 & 1.6081 & 12.09 & 0.1648 \\
$a=0.324, \rho_0=3.0563,$   & 1.5  & 1.805  & 12.41 & 0.1784\\
$n_0=0.19611 $                       & 1.75 & 2.1434 & 12.84 & 0.2011 \\
\hline
                       & & & &  \\
SQSB60                 & 0.5   &  0.5899 & 8.026 & 0.0920\\
                       &1      & 1.2296 &   9.877& 0.1495\\
$a=1/3, \rho_0=4.2785,$   &1.35   & 1.7076  &  10.68& 0.1867\\
    $n_0=0.28665$                      &1.65   & 2.1406 &   11.10& 0.2196 \\
\hline
                        & & & & \\
SQSB56                      &0.5    & 0.5383 & 7.865 & 0.0939\\
                            &0.7   &  0.7668 &  8.709& 0.1187\\
$a=0.301, \rho_0=4.4997,$  &1     &  1.1233 &  9.637& 0.1532\\
  $n_0=0.27472$             &1.2   &  1.3707 &  10.09& 0.1756\\
                       &1.35  &  1.5620 &  10.35& 0.1925\\
                      &1.5   &  1.7587 &  10.54& 0.2101\\
                      &1.65  &  1.9617 &  10.62 &0.2295\\
\hline
                      & & & &\\
DSQS                  & 1.35& 1.7191 &  7.336 & 0.2717 \\
$a=0.463, \rho_0=11.53, $ & & & &\\
   $n_0=0.725$   & & & & \\
\hline
\end{tabular}
\end{center}
\caption{Global parameters of isolated static strange stars for the
  four models of strange stars used in our computations. The symbols
  have following meaning: $M$ is the gravitational mass, $M_{\rm B}$
  the baryon mass, $R$ stellar radius, $M/R\equiv GM/Rc^2$ is the
  compaction parameter. The mass density $\rho_0$ and baryon density $n_0$ at zero pressure are in
  units $[10^{14}\rm g/ cm^3]$ and [$\rm fm^{-3}$] respectively.}
\label{EOStable}
\end{table}

\begin{figure}
\begin{center}
\includegraphics[angle=0,width=0.5\textwidth]{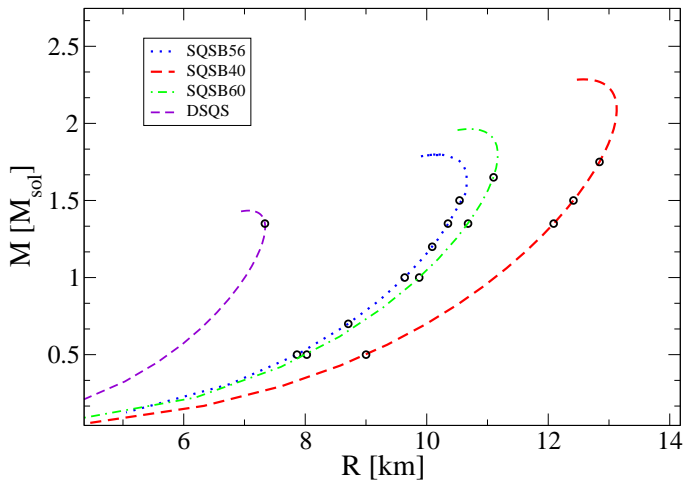}
\caption{Gravitational mass $M$ versus stellar radius $R$ for
sequences of static isolated strange quark stars described by three different
sets of parameters of the MIT-bag model and by the Dey et al. (1998)
model. The circles correspond to stellar configurations considered in
calculations of equal-mass evolutionary sequences.
} 
\label{fig1}                                 
\end{center}                                
\end{figure}

Strange quark stars (SQS) are currently considered as a possible alternative to
neutron stars as compact objects (see e.g. \cite{Weber04, Madsen99, Gondek03}
and references therein).  Typically, strange stars are modeled with an EOS
based on the MIT-bag model (e.g. \cite{AlcocFO86, HaensZS86}) in which quark
confinement is described by an energy term proportional to the volume
\cite{FahriJ84}. There are three physical quantities entering the MIT-bag
model: the mass of the strange quarks, $m_{\rm s}$, the bag constant, $B$, and
the strength of the QCD coupling constant $\alpha$. In the framework of this
model the quark matter is composed of massless u and d quarks, massive s quarks
and electrons.  We performed calculations for three different sets of
parameters of the MIT-bag model:

i) {\bf SQSB56} - the standard MIT bag model: $m_{\rm s}c^2=200\ {\rm MeV}$,
$\alpha=0.2$, $B=56~{\rm MeV/fm^3}$;

ii) {\bf SQSB60} -  the simplified MIT bag model with  $m_{\rm s}=0$, $\alpha=0$;
$B=60\ {\rm MeV/fm^3}$;

iii) {\bf SQSB40} - the ''extreme'' MIT bag model (relatively low strange 
quark mass and $B$ but high $\alpha$) : $m_{\rm s}c^2=100\ {\rm MeV}$,
$\alpha=0.6$, $B=40\ {\rm MeV/fm^3}$.

The second type of EOS which we employ is the Dey et al. (1998)
EOS of strange quark matter. In this model, quarks of the density dependent
mass are confined at zero pressure and deconfined at high density. The quark
interaction is described by an interquark vector potential originating from
gluon exchange, and by a density dependent scalar potential which restores the
chiral symmetry at high densities.  This model, with an appropriate choice of
the EOS parameters, gives absolutely stable strange quark matter. Two cases of
this model have been used in the literature SS1 and SS2 - both giving a rather
low value for the maximum gravitational mass $M_{\rm max}=1.33\,M_\odot$ and $
M_{\rm max}=1.44\,M_\odot $ respectively. We have chosen the SS2 model and 
call it
{\bf DSQS} in our paper.  The stars described by the Dey et al. (1998) model 
are very
compact i.e.  the gravitational redshifts $z$ for the maximum mass
configurations are much larger than those for strange stars  
within the MIT bag model
(also larger than $z$ for most models of neutron stars).

It was shown that different strange quark EOS can be fitted very well
by following formulas  \cite{Gondek00, Zdunik00}:

\beq \label{e:sseos} p=a(\rho-\rho_0), 
\eeq \beq \label{sseosn} n(p) =n_0\cdot\left[1+{1+a\over
    a}{p\over\rho_0}\right]^{1/(1+a)}, 
\eeq

where $p,\ \rho,\ n$ are the pressure, the mass density and the baryon density
respectively and $a,\ \rho_0,\ n_0$ are some constants.  In general
this equation corresponds to a self-bound matter with mass density
$\rho_0$ and baryon density $n_0$ at zero pressure and with a fixed
sound velocity $\sqrt{a}$ at all pressures.  The parameters $a$ and
$\rho_0$ and $n_0$ for each EOS used in the paper are given in Table I.  The
strange stars described by the DSQS model have very high density at
the surface $\rho_0=1.15\times 10^{15}~[{\rm g/cm^{3}}]$.  For the MIT
bag model $\rho_0$ is in the range $\sim 3 - 6.4 \ [ 10^{14} {\rm
    g/cm}^3] $ \cite{Zdunik00, Gondek03}.  The parameter $a$ is found
to be between 0.289 and 1/3 (for $0 \le \alpha \le 0.6 \ {\rm and}\ 0
\le m_{\rm s} \le 250 \ {\rm MeV}$) \cite{Zdunik00} for the MIT bag
model and 0.463 for DSQS model \cite{Gondek00}. The higher value of
$a$ and $\rho_0$ the higher compactness parameter of a star with fixed
gravitational mass.

In Fig. 1 we present gravitational mass versus stellar radius for sequences of
static strange quark stars. Circles correspond to configurations studied in
the paper. The global parameters are given in Table I.  Depending on the model
we obtain the radius of a 1.35 $M_\odot$ star in the range 7-13 km.

\section{Assumptions and Methods}
\subsection{Assumptions}

The first assumption regards the matter stress-energy tensor
$\w{T}$, which we
assume to have the {\it perfect fluid} form: 
\beq 
    \w{T} = (e+p)\w{u}\otimes\w{u} +p \, \w{g}, 
\eeq 
where $e$, $p$, $\w{u}$ and $\w{g}$ are respectively the fluid proper energy 
density, the fluid pressure, the fluid 4-velocity, and the spacetime metric. 
This constitutes an excellent approximation for neutron star matter or 
strange star matter.

The last orbits of inspiraling binary compact stars can be studied in
the {\it quasi-equilibrium} approximation. Under this assumption the
evolution of a system is approximated by a sequence of exactly
circular orbits. This assumption results from the fact that the time
evolution of an orbit is still much larger than the orbital period and that 
the gravitational radiation circularizes an orbit of a binary system.
This implies a continuous spacetime symmetry, called  {\em helical
symmetry} \cite{BonazGM97,FriedUS02}  represented by the Killing
vector 
\beq \w{\ell} = {\p \over \p t} +\Omega {\p \over \p \vp}, \eeq
where $\Omega$ is the orbital angular velocity and $\p/\p t$ and
$\p/\p \vp$ are the natural frame vectors associated with the time
coordinate $t$ and the azimuthal coordinate $\vp$ of an asymptotic
inertial observer. 

We also assume that the spatial part of the metric 
(i.e. the metric induced by $\w{g}$ on each hypersurface $\Sigma_t$)
is conformally flat,
which corresponds to the {\it Isenberg-Wilson-Mathews}
approximation to general relativity \cite{Isenb78,IsenbN80,WilsoM89}
(see Ref.  \cite{FriedUS02} for a discussion). Thanks to this
approximation we have to solve only five of the ten Einstein
equations. 

The fourth assumption concerns the fluid motion inside each star.  We
only consider {\it irrotational} motion  (assuming that
the fluid has zero vorticity in the inertial frame). 

\subsection{Equations to be solved}

We just mention briefly all the equations we have to solve and 
refer the reader to Limousin et al. \cite{LimouGG05} for more details. 

The gravitational field equations have been obtained within the 3+1 
decomposition of the Einstein's equations \cite{York79,Cook00}, using the
extended conformal thin sandwich formalism \cite{PfeifY03} and taking 
into account the helical symmetry of the spacetime. 
This gives one vectorial elliptic
equation for the shift vector $N^i$ coming from the momentum constraint and two scalar elliptic equations for $\nu = {\rm ln}N$ and
$\beta={\rm ln}(AN)$, coming from the trace of the spatial part of the 
Einstein equations combined with the Hamiltonian constraint, $N$ being the 
lapse function and $A$ the conformal factor. \\

Apart from the gravitational field equations, we have to solve the
fluid equations.  The equations governing the quasi-equilibrium state
are the relativistic Euler equation and the equation of baryon number 
conservation. Irrotational motion admit a first
integral of the relativistic Euler equation:
\beq 
H + \nu - \ln \Gamma_0 + \ln \Gamma = {\rm const.}, 
\eeq
where $H$ is the pseudoenthalpy, $\Gamma_0$ is the Lorentz factor between
the co-orbiting  and
the Eulerian observers and $\Gamma$ is the Lorentz factor between the fluid 
and the co-orbiting observers. \\

The equation of baryon number conservation is written as an elliptic equation 
for the velocity potential $\Psi$. The method of solving this equation is 
different
for neutron stars and strange stars. For strange stars, we have to impose 
a Neumann-like boundary condition for the velocity potential at the surface of 
the star (see paragraph IV.C of \cite{LimouGG05} for details) to have flow field tangent to the surface in a rotating frame.

\subsection{Numerical method}

The resolution of the above nonlinear elliptic equations is performed thanks
to a numerical code based on multidomain spectral methods and constructed upon
the {\it LORENE} C++ library \cite{Lorene}.  The detailed description of the
whole algorithm, as well as numerous tests of the code can be found in
\cite{GourGTMB01}.  Additional tests have been presented in Sec.~3 of
\cite{TanigG03}.  The code has already been used successfully for calculating
the final phase of inspiral of binary neutron stars described by polytropic
EOS \cite{Faber02,BonazGM99,TanigGB01,TanigG02a,TanigG02b,TanigG03}, realistic EOS
\cite{Gondek07,Bejger05} as well as binary strange stars \cite{Gondek07,LimouGG05}.  It is worth
to stress that the adaptation of the domains (numerical grids) to the stellar
surface (surface-fitted coordinates) used in the code is very important for
calculating binary systems of strange stars. This method enable us to treat
the strongly discontinuous density profile at the surface of a strange star
and avoid any Gibbs-like phenomenon.
\cite{BonazGM98}.

We used one numerical domain for each star and 
3 (resp. 4) domains for the space around
them for a small (resp. large) separation between stars. In each domain, 
the number
of collocation points is chosen to be $N_r
\times N_{\theta} \times N_{\varphi} = 25 \times 17 \times 16$, where
$N_r$, $N_{\theta}$, and $N_{\varphi}$ denote the number of 
collocation points ($=$ number of polynomials used in the spectral
method) in
the radial, polar, and azimuthal directions respectively.  

The convergence of the procedure is monitored by computing the relative
difference $\delta H/H$ between the enthalpy fields at two successive
steps. The iterative procedure is stopped when $\delta H/H$ goes below a
certain threshold, typically $10^{-7}$.  The accuracy of the computed models
is estimated using a relativistic generalization of the virial theorem
\cite{FriedUS02}. The virial relative error was found to be a few times
$10^{-5}$ for the closest configurations.

\section{Evolutionary sequences}

The evolution of a binary system of compact objects is entirely driven by
gravitational radiation and can be roughly divided into three phases :
point-like inspiral, {\it hydrostationary inspiral} and merger. The first phase
corresponds to large orbital separation (much larger than the neutron star
radius) and can be treated analytically using the post-Newtonian (PN)
approximation to general relativity (see Ref.~\cite{Blanc02b,Blanc06} for a
review). In the second phase the orbital separation becomes only a few times
larger than the radius of the star, so the effects of tidal deformation,
finite size and hydrodynamics play an important role.  In this phase, since
the shrinking time of the orbital radius due to the emission of gravitational
waves is still larger than the orbital period, it is possible to approximate
the state as quasi-equilibrium \cite{BaumgCSST97,BonazGM99}.  The final phase
of the evolution is the merger of the two objects, which occur at the
dynamical timescale
\cite{ShibaU00,ShibaU01,ShibaTU03,OoharN99}. The quasi-equilibrium
computations from the second phase provide valuable initial data for the
merger \cite{ShibaU00,ShibaTU03,Oechslin04,FaberGR04}.

We focus on the last orbits of the inspiral phase (the hydrostationary
inspiral).  In this section, we present the numerical results for evolutionary
sequences of close strange stars binaries described by three different sets of
parameters of the MIT bag model and the Dey model introduced in Sect. II. We
consider only equal-mass binary systems with different total masses. By
evolutionary sequence, we mean a sequence of quasi-equilibrium configurations
of decreasing separation and with constant baryon mass $M_{\rm B}$, which are
expected to approximate the true evolution of a binary system.  In order to
investigate the properties of the GW emission during the final phase of binary
strange star inspiral we focus on the variation of the ADM
(Arnowitt-Deser-Misner, see e.g.\cite{TanigG03}) mass of the system $M_{\rm
ADM}$ (the total binary mass-energy) with respect to the GW frequency. These
two quantities are sufficient to determine the GW energy spectrum.  The
orbital binding energy is defined by

\beq  E_{{\rm bind}} := M_{{\rm ADM}} - M_\infty, \eeq 

where $M_\infty$ is the ADM mass of the system at infinite
separation, that is the sum of the gravitational masses, $M$, of isolated static
stars. The variation of $E_{{\rm bind}}$ along an
evolutionary sequence corresponds to the loss of energy via gravitational
radiation. Gravitational waves are emitted at twice the orbital
frequency: $f_{{\rm GW}} = 2 f = \Omega / \pi$. \\

The physical inspiral of binary compact stars terminates by either the orbital
instability (turning point of $E_{\rm bind}$) or the mass-shedding limit (when
a cusp forms at the stellar surface in the direction of the companion (Roche
lobe overflow)). In both cases, this defines the {\it innermost stable
circular orbit}.  The frequency of gravitational waves at the ISCO is one of
potentially observable parameters by the gravitational wave detectors.
In addition so called ``{\it break frequencies}'', characteristic frequencies where
the power emitted in gravitational waves decreases measurably could be observable quantities \cite{Faber02,Hughes02} and section V. C at this paper.

\section{Results for equal-mass strange star binaries with $M_\infty = 2.7 \, M_\odot$} 
\begin{figure}{}
\begin{center}
\includegraphics[angle=-90,width=0.5\textwidth]{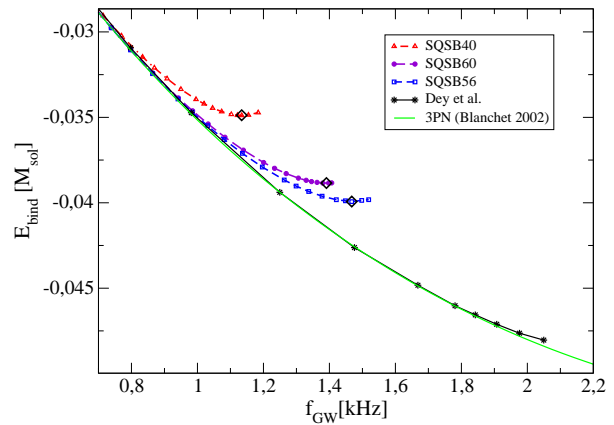}
\caption{ Orbital binding energy  $E_{{\rm bind}}=M_{ADM}-2M$ as a function of 
gravitational wave frequency (twice the orbital frequency) along evolutionary
sequences of irrotational equal mass (of 1.35 $M_{\odot}$) strange star
binaries described by the four EOS introduced in 
Section II. The solid line corresponds to the 3rd post-Newtonian point masses
approximation  derived by Blanchet \cite{Blanc02}.}
\label{f:E_f}
\end{center}
\end{figure}

In Fig. \ref{f:E_f} we present the orbital binding energy as a function of
gravitational wave frequency along evolutionary sequences of strange quark
stars with the total mass-energy in infinite separation $M_\infty = 2\times
1.35 \ M_\odot = 2.7 \, M_\odot$.  The different symbols (triangles, stars, diamonds and
squares) indicate the individual equilibrium configurations calculated
numerically. The big diamonds correspond to the minimum of the binding energy
of an evolutionary sequence. A turning point of $E_{{\rm bind}}$ along an
evolutionary sequence indicate an orbital instability \cite{FriedUS02}. This
instability can originate both from relativistic effects (the well-known $r=6M$
last stable orbit of Schwarzschild metric) and hydrodynamical effects \cite{UryuSE00}.

We present also the 3rd order point masses post-Newtonian (PN)
approximation derived by Blanchet \cite{Blanc02} (solid
line). Comparison of our numerical results with the 3PN calculations
reveals a good agreement for small frequencies (large separations).
The deviation from PN curves at higher frequencies (smaller
separation) is due to hydrodynamical effects, which are not taken into
account in the PN approach. The relative difference between binding
energy of quasiequilibrium sequences of two $1.35 M_\odot$ strange
stars and the 3PN point-mass calculation, caused by the finite size
effects, at the frequency of gravitational waves corresponding to the
quasiequilibrium ISCO is $\sim 7 \%$.

\begin{figure}{}
\begin{center}
\includegraphics[angle=-90,clip,width=0.45\textwidth]{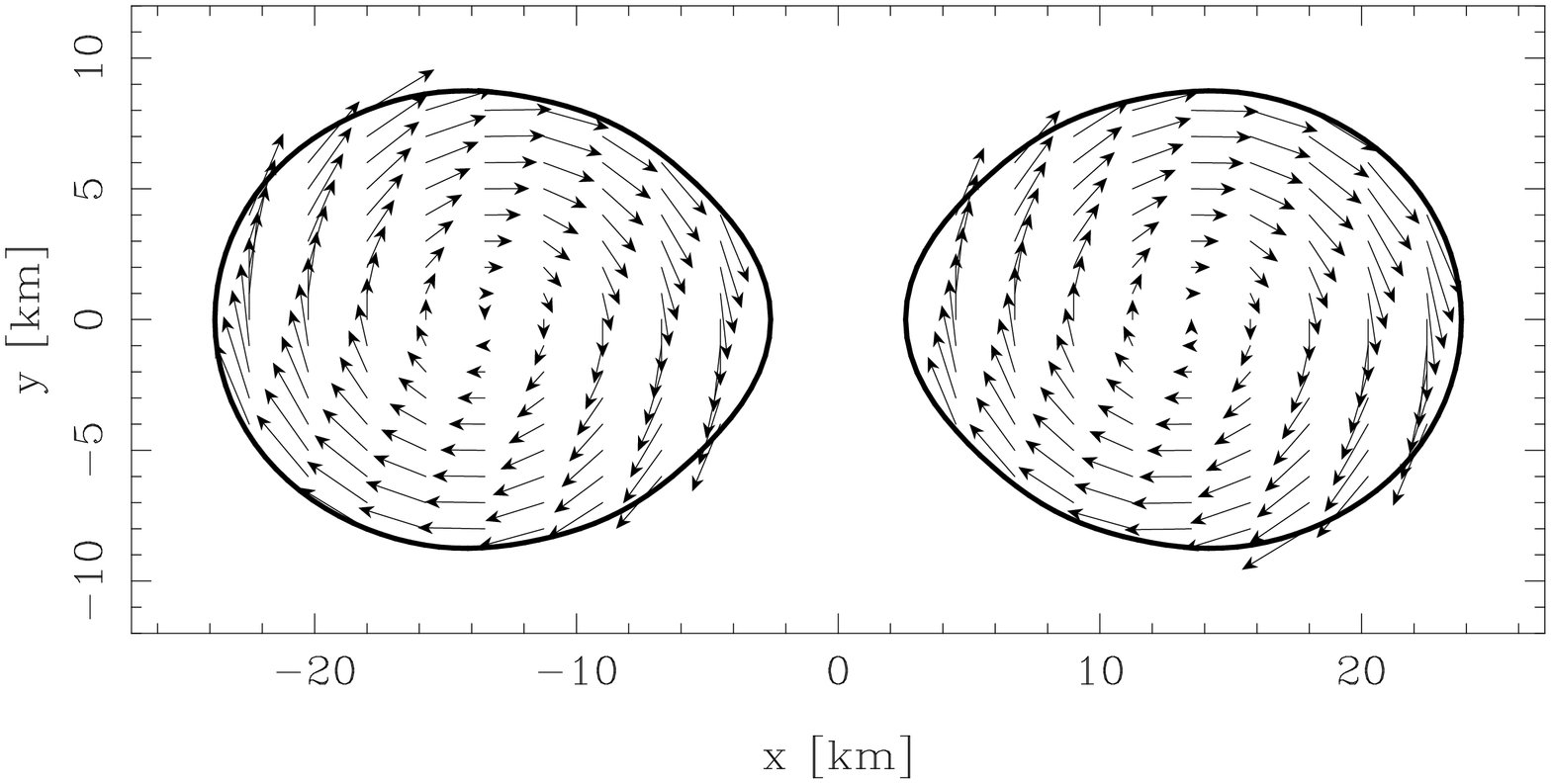}
\includegraphics[angle=-90,clip,width=0.45\textwidth]{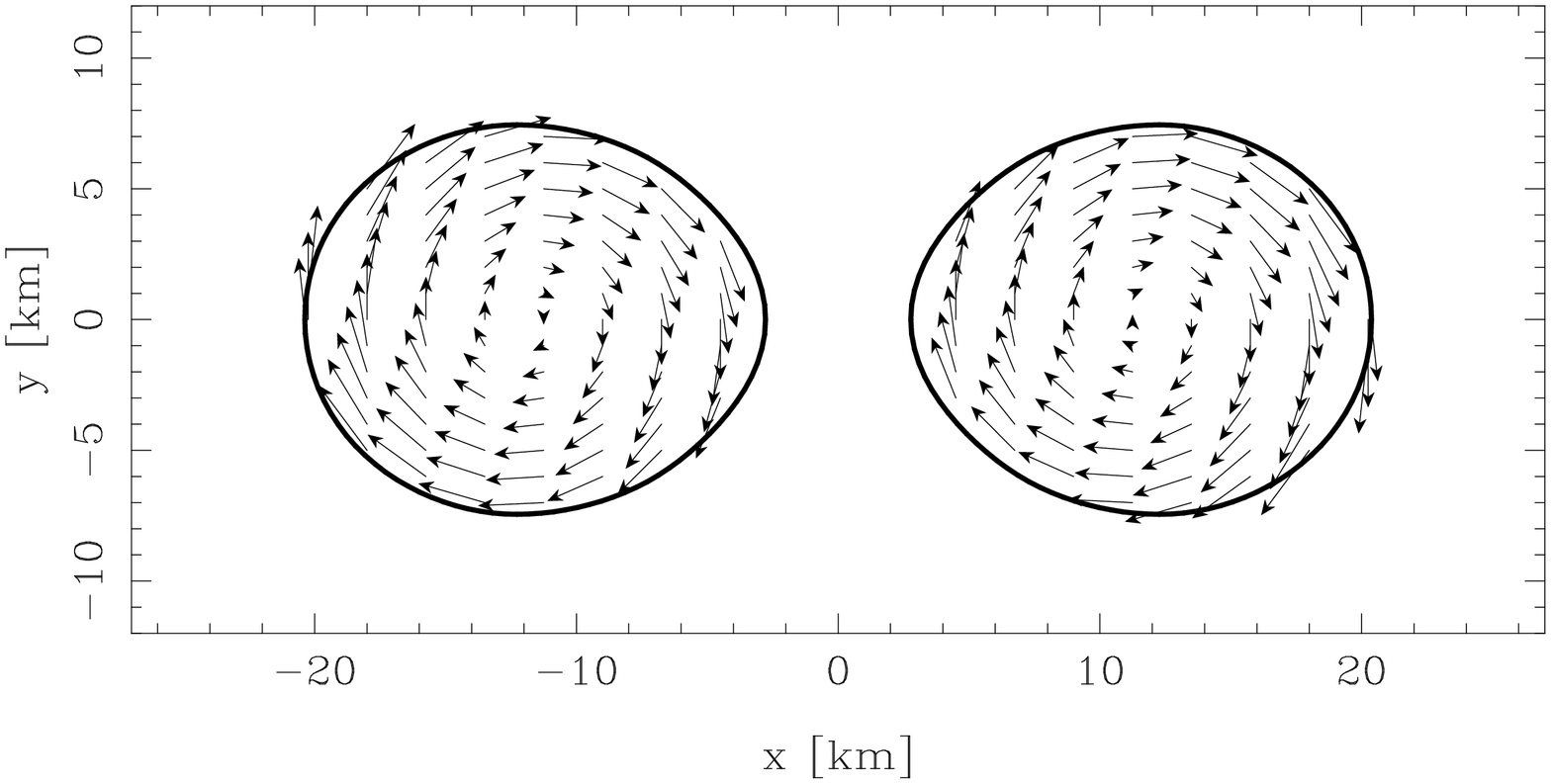}
\includegraphics[angle=-90,clip,width=0.45\textwidth]{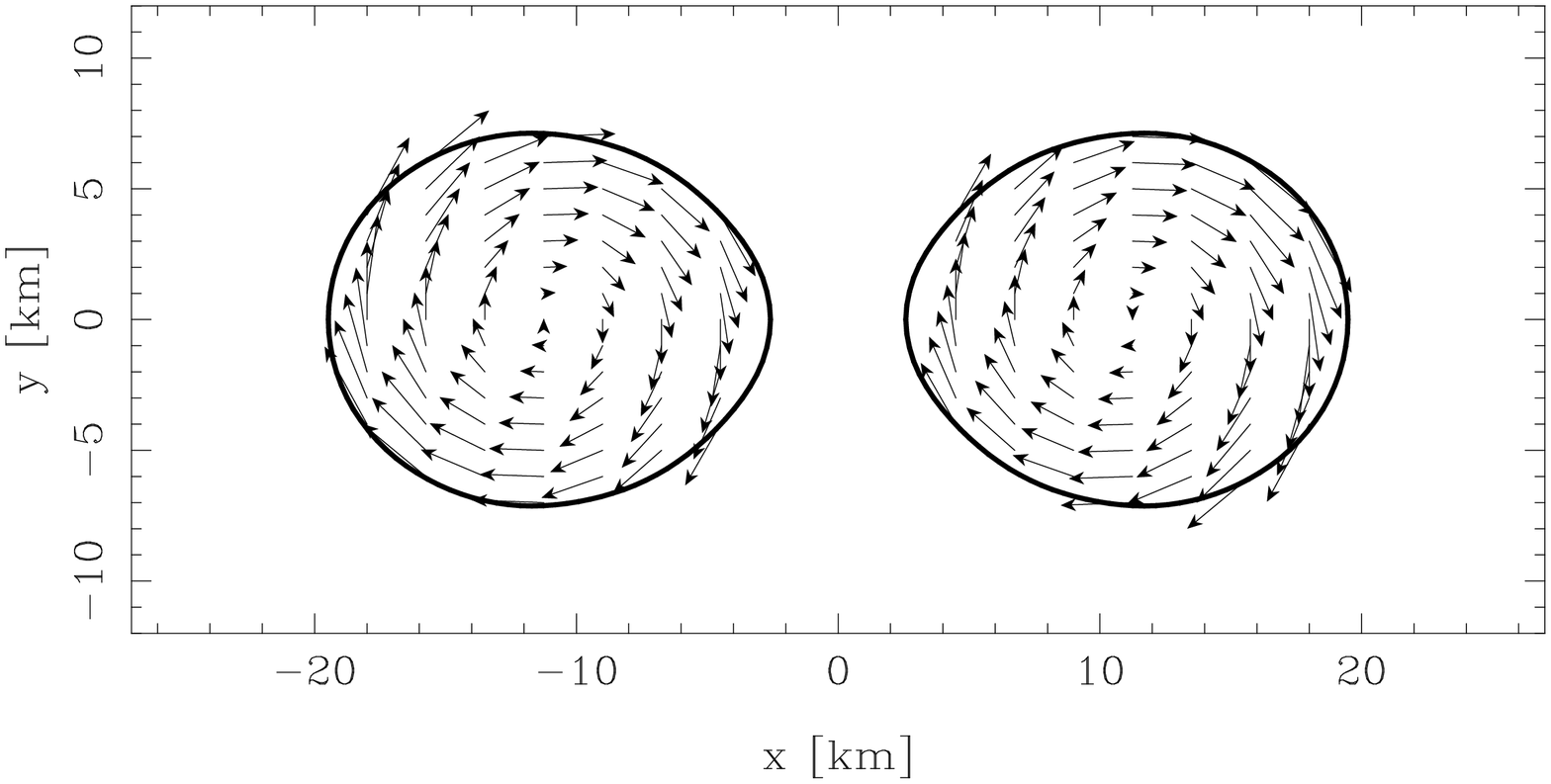}
\includegraphics[angle=-90,width=0.45\textwidth]{Dey_velocity.ps}
\caption{Velocity field with respect to the co-orbiting frame in the orbital
 plane for strange stars at the coordinate separation corresponding to ISCO
 (three upper panels) or to the closest calculated configuration (the lowest
 panel). The panels correspond to three types of the MIT bag model, SQSB40,
 SQSB60, SQSB56 and Dey et al. (1998) EOS SQSD from the upper to the lower one
 respectively.  The thick solid lines denote stellar surfaces.}
\label{f:velocity}
\end{center}
\end{figure}

A turning point of $E_{{\rm bind}}$ is found for each of the three binary
strange stars described by the MIT bag model but by the DSQS.  For the DSQS
model we had to finish our calculations before the orbital instability was
reached since for closer configurations we were not able to obtained required
accuracy due to very high compactness parameter for this strange quark model.
The shape of the MIT bag model strange stars at the ISCO and the closest
computed configuration in the case of DSQS model are presented in
Fig. \ref{f:velocity}. The stars for DSQS model are in fact still nearly
spherical. One can thus suppose that the turning point should exist for
the DSQS model of strange matter in high frequencies.
The frequency of gravitational waves at the
ISCO for the MIT bag model is found to be in the range $\sim 1130 -1470$Hz .
The 3PN results for point masses derived by
different authors give ISCO at very high frequencies of gravitational waves $>
2\ {\rm kHz}$ \cite{Blanc02, DamouJS00, DamouGG02}

\subsection{The impact of EOS on the GW frequency at ISCO}

\begin{figure}{}
\includegraphics[angle=-90,width=0.5\textwidth]{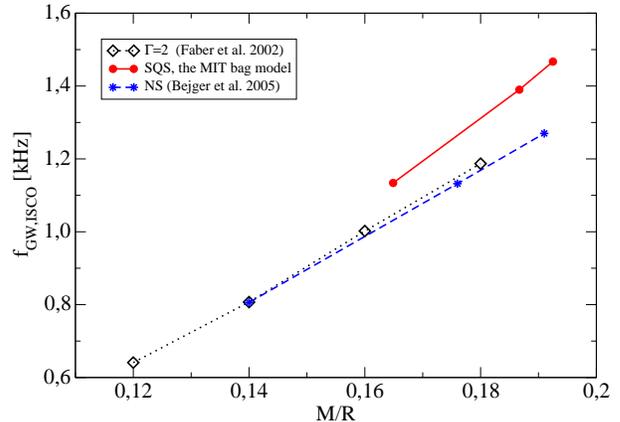}
\vskip 0.5cm
\caption{The frequency of ISCO (the dynamical orbital instability for strange
  quark stars or the mass-shedding limit for neutron stars) versus compaction
  parameter for equal-mass (of $1.35 M_\odot$) strange stars and neutron stars
  described by realistic EOS \cite{Bejger05} and polytropic EOS \cite{Faber02}
  }
\label{f:f_MR}
\end{figure}

It was already suggested by many authors that the frequency of GW at the ISCO
depends on the compactness parameter and thus on EOS of nuclear
matter.  In Fig 4. we present the frequency of GW at ISCO as a function of
compactness parameter for equal-mass binary (consisting of two $1.35 M_\odot$)
strange stars described by the MIT bag model and neutron stars described by
three nuclear EOS (GNH3 EOS \cite{Glend1985}, APR EOS \cite{akmal},
BPAL12 EOS \cite{PrakaBPELK97,Bombaci95}- see \cite{Bejger05} for details) and
by polytropic EOS with $\gamma = 2$ \cite{Faber02}.

All evolutionary sequences of binary strange stars terminate at the dynamical
orbital instability while equilibrium sequences of binary neutron stars at the
mass-shedding limit (for polytropic EOS with $\gamma \le 2.5$ the ISCO is
given by the mass-shedding limit otherwise by the orbital instability
\cite{UryuSE00, TanigG03}). For all EOS the higher compactness of a star is,
the higher frequency of gravitational waves at the ISCO is.

We see that for equal-mass evolutionary sequences with the same total mass the
dependence $f_{\rm GW,ISCO}$ versus the compactness parameter can be described
by linear function $y=A1x+A2 $ for all EOS, where A1 is found to be 11.97,
9.09, 9.165 and A2: -0.841, -0.469,-0.466 for SS binaries, NS described by
realistic EOS and polytropic NS binaries respectively. We see that the results
obtained for polytropic EOS fit quite well the calculations performed for
neutron stars described by realistic EOS.  Indeed as found by Bejger et
al. 2005 \cite{Bejger05} the frequency of gravitational waves at the end point
of inspiraling neutron stars described by several realistic EOS without exotic
phases (such as meson condensates or quark matter) can be predicted, in a good
approximation, by studying binaries with assumed polytropic EOSs with
$\gamma=2$ or 2.5.  As found by Limousin et al. \cite{LimouGG05} it wasn't the
case for inspiraling strange star binaries which are self-bound objects having
very large adiabatic index in the outer layers. The frequency of gravitational
waves at the end of inspiral phase is higher by $\sim 150$ Hz for irrotational
strange star binaries than for the polytropic neutron star binaries with the
same gravitational mass and stellar radius in infinite separation. The
differences in the evolution of binary strange stars and neutron stars stem
from the fact that strange stars are principally bound by another force than
gravitation: the strong interaction between quarks. Thanks to this, at the end
of the inspiral phase, neutron stars are, for the same separation, more oblate
than strange stars. And thus a cusp forms at the stellar surface of neutron
stars, which marks the beginning of exchange of matter between the two stars,
whereas the surface of strange stars is smooth even at the dynamical
instability (see Fig. \ref{f:velocity}).

The frequency at the end point of inspiral of binary neutron stars described
by three nuclear EOS ranges from $806$ Hz for GNH3 EOS ($M/R=0.14$) to $1270$
Hz for BPAL 12 EOS ($M/R = 0.191$).  The range of frequencies at the ISCO for
binary neutron stars intersects with these for binary strange stars (higher
than 1130 Hz).  The determination of the gravitational wave frequency at the
ISCO by the laser interferometers allows to impose constraints on the EOS of
matter at ultra-high densities but it is not sufficient to distinguish
completely between strange stars and neutron stars. The detection of high
value of $f_{\rm GW,ISCO}$ e.g. 1250 Hz from binary system consisting of two
1.35 $M_\odot$ stars could indicate the compactness parameter 0.175 or 0.189
depending on EOS.   \\

\subsection{Analytical fits to numerical results}

As already mentioned the 3PN calculations reproduce quite well our
results for small frequencies (large separations). For close separation we see
the deviation from  point-mass calculations due to
hydrodynamical effects.  The observed deviation of the GW energy spectrum
for quasiequilibrium sequences (given by the derivative $dE_{\rm bind}/df_{\rm
  GW}$) from point-mass behavior gives also important information about EOS
of neutron stars in addition to the frequency of GW at the ISCO
\cite{Faber02}. 

Following Faber et. al. (2002) \cite{Faber02} and Bejger et. al. (2005)
\cite{Bejger05}, we perform some polynomial fits (see below) to each of the
computed evolutionary sequences in order to obtain functions required for the
GW energy spectrum. The two different approaches were used by authors to
represent the variation of the total mass energy as a function of GW
frequency.  Faber et. al. (2002) \cite{Faber02} fitted the numerical results
taking into account 3 terms: $f^{2/3},\ f\ {\rm and}\  f^2$ representing the
Newtonian point-mass behavior, the lowest order post-Newtonian and
finite-size corrections, the tidal interaction energy respectively.
However Bejger et. al. (2005) \cite{Bejger05} found that it is possible to
find much better approximations of numerical results taking into account higher
order PN terms. They have shown  that the difference between the
binding energy of equal-mass of 1.35 $M_{\odot}$ irrotational neutron star
binaries and the binding energy of binary point masses in the 3PN
approximation of Blanchet (2002) \cite{Blanc02} 
can be fitted very well by the power-law dependence on frequency
$f_{{\rm GW}}$: \beq E_{{\rm bind}} - E_{{\rm bind}}^{{\rm 3PN}} = A
\left(\frac{f_{{\rm GW}}} {1000 {\rm Hz}}\right) ^n.
\label{e:fit}
\eeq

The 3PN formula as obtained by Blanchet
\cite{Blanc02} from the standard post-Newtonian expansion reads \bea
\frac{E_{{\rm bind}}^{{\rm 3PN}}}{M_\infty} = &-& \frac{1}{8} \Omega_{*}^{2/3}
+ \frac{37}{384}\Omega_{*}^{4/3} + \frac{1069}{3072}\Omega_{*}^2 \nonumber \\
&+& \frac{5}{3072}\left(41\pi^2 - \frac{285473}{864}\right) \Omega_{*}^{8/3},
\label{e:Ebind_3PN}
\eea
where $\Omega_{*}$ is the orbital angular frequency expressed in geometrized 
units:
\beq
\Omega_{*} := 2\pi M_\infty f = 2\pi M  f_{{\rm GW}} = 2M\Omega.
\eeq
The terms in $\Omega_{*}^{2/3}$, $\Omega_{*}^{4/3}$, $\Omega_{*}^2$ and 
$\Omega_{*}^{8/3}$ in Eq. (\ref{e:Ebind_3PN}) are respectively the
Newtonian, 1PN, 2PN and 3PN term.

\begin{figure}{}
\includegraphics[angle=-90,width=0.5\textwidth]{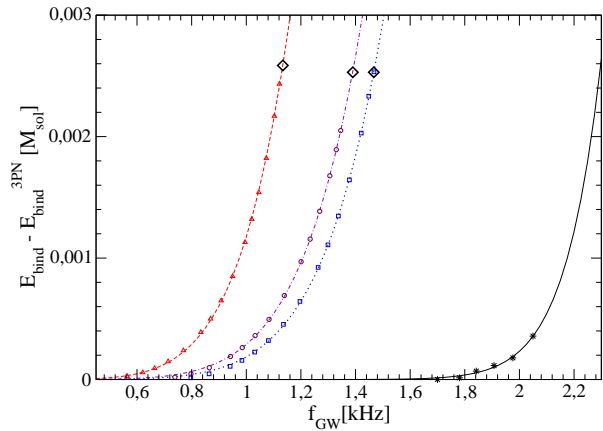}
\caption{Difference $E_{\rm bind} - E_{\rm bind}^{3PN} $ between the binding
 energy of equal-mass of 1.35 $M_{\odot}$ irrotational strange quark star
 binaries and the binding energy of binary point masses in the 3PN
 approximation of Blanchet (2002) \cite{Blanc02}.  The different symbols
 (triangles, circles, squares and stars) correspond to the numerical results, lines to polynomial fits (\ref{e:fit}) to them and big diamonds to ISCO.}
\label{f:E_E3PN}
\end{figure}

In Fig. \ref{f:E_E3PN}, we present the difference between our numerical 
results and the 3PN approximation given by Eq. (\ref{e:Ebind_3PN}).
Looking at the scale of Fig. \ref{f:E_E3PN}, we see that the formula 
(\ref{e:Ebind_3PN}) approximates very well the behavior of a binary
system of strange stars for a large range of frequencies.  

Because of the steep character of the function 
$E_{{\rm bind}} - E_{{\rm bind}}^{{\rm 3PN}}$ seen in Fig. \ref{f:E_E3PN},
the power $n$ is quite large. The values are listed in Table \ref{fittable}.
We didn't assume the integer number of the power $n$. 
We note that the values of the power $n$ are similar for the three EOS of 
MIT bag model. So we also done fits for an intermediate value $n=6.5$ for these
three EOS and denote by $A_{n=6.5}$ its corresponding factor. 
The information on the frequency of departure from 3PN 
curve is thus entirely determined by the numerical factor $A$. The higher this 
factor is,
the higher the frequency of departure. For Dey et. al. (1998) EOS, the power
$n$ is very high and the reference frequency of $1000$ Hz used for the 
polynomial fit (\ref{e:fit}) is not well adapted to the frequency of 
gravitational waves obtained using this EOS, explaining the very small
factor $A$.

From Fig.
\ref{f:E_E3PN}, we can define the frequencies $f_{{\rm npm}}$ as those
frequencies at which the deviation from point-mass behavior becomes 
important. It can be defined more precisely 
by the frequency for which
\beq
\frac{E_{{\rm bind}} - E_{{\rm bind}}^{{\rm 3PN}}}
{E_{{\rm bind}}^{{\rm 3PN}}} = 0.001.
\eeq
The values of these frequencies for the four strange star models
are given in Table \ref{ftable}. 

\begin{table}[h]
\begin{center}
\begin{tabular}{|c|c|c|c|c|}
\hline
EOS &$M/R$ & $A~[M_{\odot}]$ & $n$ & $A_{n=6.5}[M_\odot]$\\
\hline\hline
$\ $SQSB40$\ $ & $\ $0.1648$\ $ & 0.001174 &6.28 & 0.001158 \\
\hline
SQSB60  & 0.1867 & 0.0002888 &6.59 & 0.0002961 \\
\hline
SQSB56 & 0.1925 & 0.0001865 &6.8 & 0.0002068\\
\hline
DSQS & 0.2717 & 1.353e-9 &17.38 & -- \\
\hline
\end{tabular}
\end{center}
\caption{Parameters $A$, $n$ and $A_{n=6.5}$ ($A$ for $n=6.5$) of polynomial
  fits (\ref{e:fit}) for equal-mass of 1.35 $M_{\odot}$ irrotational strange
  quark star binaries.}
\label{fittable}
\end{table}

One can draw an important conclusion from the presented results and their 
comparison with relativistic approximations for point masses in a binary
system. We can expect that taking into account the next orders in a 
post-Newtonian approximation doesn't change the energy by an amount larger 
than the difference between 2PN and 3PN models. As a consequence the large
deviation of our numerical results from the 3PN  approximation is caused 
by the effects of a finite size of the star, e.g. tidal forces. The very high
power $n$ indicates that high order tidal effects are very important, and 
dominates the relation $E_{{\rm bind}}(f_{{\rm GW}})$. Indeed, the lowest
order tidal term is known to be $n=4$ \cite{LaiRS94} and the values 
obtained here are well above this. 

\subsection{Energy spectrum of gravitational waves}

\begin{figure}{}
\includegraphics[angle=-90,width=0.5\textwidth]{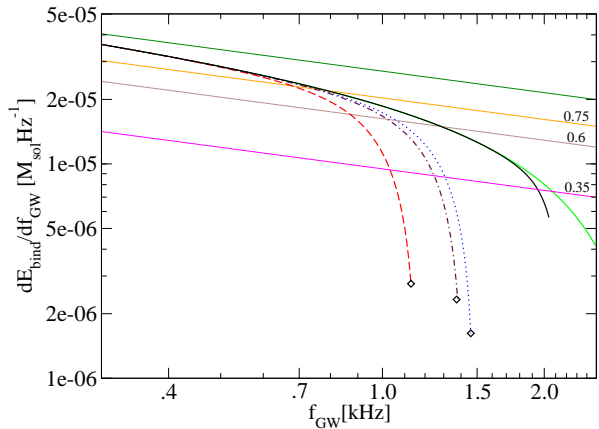}
\caption{Energy spectrum of gravitational waves emitted by strange stars
binaries versus frequency of gravitational waves along the four equal-mass of
1.35 $M_{\odot}$ irrotational quasiequilibrium sequences. The straight lines
correspond to the Newtonian dependence of energy multiplied by 1, 0.75, 0.6
and 0.35.}
\label{f:spectrum}
\end{figure}

\begin{table}[h]
\begin{center}
\begin{tabular}{|c|c|c|c|c|c|c|}
\hline
EOS & $M/R$ & $f_{\rm npm}$ & $f_{\rm 25}$& $f_{\rm 40}$ & $f_{\rm 65}$ & $f_{\rm ISCO/end}$ \\
\hline\hline
$\ $SQSB40 $\ $&  0.1648 & 542 & 679 & 871 &  1033 &1134  \\
\hline
SQSB60 & 0.1867 & 705 & 744 & 1022 & 1245 & 1390   \\
\hline
SQSB56 & 0.1925 & 767 & 756 & 1063 & 1308 &1467  \\
\hline
DSQS & 0.2717 & 1825 & 786 & 1286 & 1945 & 2050(*)\\ 
\hline
\end{tabular}
\end{center}
\caption{Gravitational wave frequencies (in Hz) at the last orbits of
  inspiraling equal-mass of 1.35 $M_{\odot}$ strange quark star
  binaries: $f_{\rm npm}$ denotes the frequency of GW at which the
  relative difference between binding energy calculated in
  quasiequilibrium and 3PN approximation is higher than $0.1 \%$,
  $f_{\rm 25},~f_{\rm 40}$ and $f_{\rm 65}$ are the so-called
  break-frequencies at which the GW energy spectrum has dropped,
  respectively, by $ 25 \%,\ 40 \%,\ 65 \% $ and $f_{\rm ISCO/end}$ is
  the GW frequency at ISCO for SQSB40, SQSB56, SQSB60 model and (*) at the
  last calculated configuration for DSQS model.}
\label{ftable}
\end{table}

\begin{figure}{}
\includegraphics[angle=-90,width=0.5\textwidth]{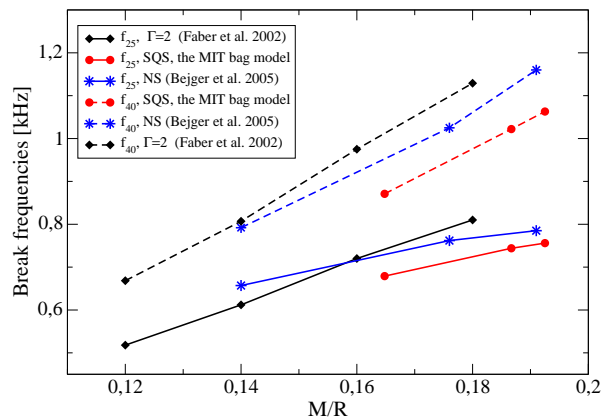}
\caption{The break frequencies  versus compaction
  parameter for equal-mass (of $1.35 M_\odot$) strange stars and neutron stars
  described by realistic EOS \cite{Bejger05} and polytropic EOS \cite{Faber02}
  }
\label{f:fbreak}
\end{figure}

We compute the energy spectrum of gravitational waves obtained as the
first derivatives of the fitted functions (\ref{e:fit}).  The quantity
${\rm d}E_{{\rm bind}}/ {\rm d}f_{\rm GW}$ is important for the data analysis
of GW, because it determines the evolution of the wave's phase.  The
difference in the phase, for two $1.35 M_\odot$ strange stars described by the MIT bag model, between
our numerical results and 3PN, caused by hydrodynamical effects, is
$\sim 40 \%$, for the last two orbits of inspiral. The phase error in the 3PN values of the wave's phase may be larger for the final orbits if one goes beyond spacial conformal flatness aproximation \cite{Uryu06}.  

The relation between 
${\rm d}E_{{\rm bind}}/ {\rm d}f_{\rm GW}$ and the gravitational waves frequency
$f_{{\rm GW}}$ is presented in Fig. \ref{f:spectrum}. In this figure, we draw
straight lines corresponding to the Newtonian case $\sim f_{{\rm GW}}^{2/3}$
to find the break frequencies $f_{25}$, $f_{40}$ and $f_{65}$ 
at which the energy spectrum has dropped respectively by
$25$ \%, $40$ \% and $65$ \%. The values of the break frequencies for the 
four EOS used to described strange stars are given in Table \ref{ftable}.
These values are important from the point of view
of future detections: they show the difference between the amplitude of the
real signal and the Newtonian template which allows to calculate the real wave
form amplitude from the detector noise. 

At the level of $25$\%, SQSB40 EOS is the only curve which deviate
from the 3PN curve so we can already distinguish SQSB40 EOS and the
other EOS. But this is only at $f_{40}$ that we can discriminate
between the four EOS, but the curve for DSQS is still very close to
the 3PN curve because of very high compaction parameter for this
model. To see the significant deviation of DSQS model from
post-Newtonian results, we need to look at frequency $f_{65}$ where
the energy spectrum has dropped by 65 \%.

In Fig. \ref{f:fbreak} we present the break frequencies $f_{25}$, $f_{40}$
for neutron stars and strange stars described by the MIT bag model versus
the compaction parameter. We find that the break frequencies
are more sensitive quantities to EOS than GW frequencies at ISCO. For
equal-mass binaries (of $1.35 M_\odot$) of neutron stars described by
realistic EOS the function $f_{\rm GW,ISCO}(M/R)$ can be quite well
reproduced studying neutron stars described by polytropic EOS with
$\Gamma =2$ (see Fig. 4). This isn't the case for the dependence of
break frequencies versus the compaction parameter.  The relation of
frequencies $f_{25}$, $f_{40}$ versus $M/R$ is different for different
EOS, and for given $M/R$ always higher for neutron stars than for
strange quark stars.

\section{The impact of the total mass and EOS on the last orbits of inspiral}

Up to now, we studied the impact of the equation of state on the
frequency of gravitational waves at ISCO and on break frequencies for
equal-mass  binaries with $M_\infty=2.7 M_\odot$. In this section, we
consider equal-mass strange star binary systems with different total
masses.

\begin{figure}{}
\vskip 1cm
\includegraphics[angle=-90,width=0.5\textwidth]{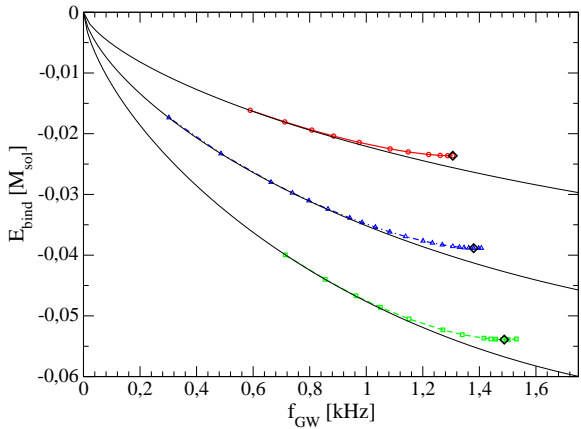}
\caption{Orbital binding energy $E_{{\rm bind}}$ as a function of GW frequency
$f_{{\rm GW}}$ along three different total mass evolutionary sequences of
irrotational binaries described by SQSB60. The total mass in infinite
separation of binaries containing two identical strange stars is 2, 2.7 and
3.3 $M_{\odot}$ from the top to the bottom respectively.
}
\label{f:E_f_diffmass}
\end{figure}

\begin{figure}{}
\vskip 1cm
\includegraphics[angle=-90,width=0.5\textwidth]{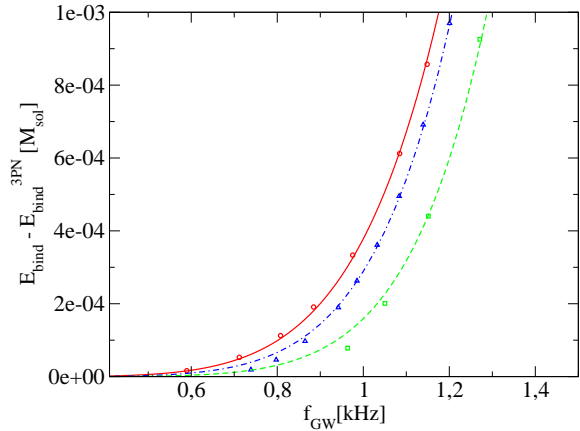}
\caption{$E_{{\rm bind}} - E_{{\rm bind}}^{3PN} $ 
for three irrotational strange star binaries described by SQSB60 model.  The symbols
correspond to the numerical results and the lines to polynomial fits to
them. The total mass of binaries containing two identical strange stars is 2,
2.7 and 3.3 $M_{\odot}$ from  left to right respectively. }
\label{f:E_E3PN_diffmass}
\end{figure}

In Fig. \ref{f:E_f_diffmass} we show the orbital binding energy versus frequency of gravitational waves along 
evolutionary sequences of binary strange stars described by SQSB60 EOS with $M_\infty= 2\ M_\odot $,\ $2.7 M_\odot$ and $3.3 M_\odot$. 

 We find a minimum of binding energy for each sequence, which
 corresponds to the dynamical instability.  The frequency of
 gravitational waves at the ISCO increases with increasing total mass
 of equal-mass irrotational strange star binaries. We find the same
 behavior for other models of strange stars.

In Fig. \ref{f:E_E3PN_diffmass} we show the difference $E_{\rm bind} -
E_{\rm bind}^{3PN} $ between the binding energy of three equal-mass
irrotational strange star binaries described by SQSB60 model and the
binding energy of binary point masses in the 3PN approximation. The
parameters of the polynomial fits Eq. (\ref{e:fit}) and $f_{\rm npm}$
frequencies for these three evolutionary sequences are given in Table
\ref{fittable_diffmass}. We find high values of the power $n$
independently on the total mass of a binary strange star system, which
indicates that tidal effects dominate the last orbits of inspiral.

\begin{figure}{}
\includegraphics[angle=-90,clip,width=0.5\textwidth]{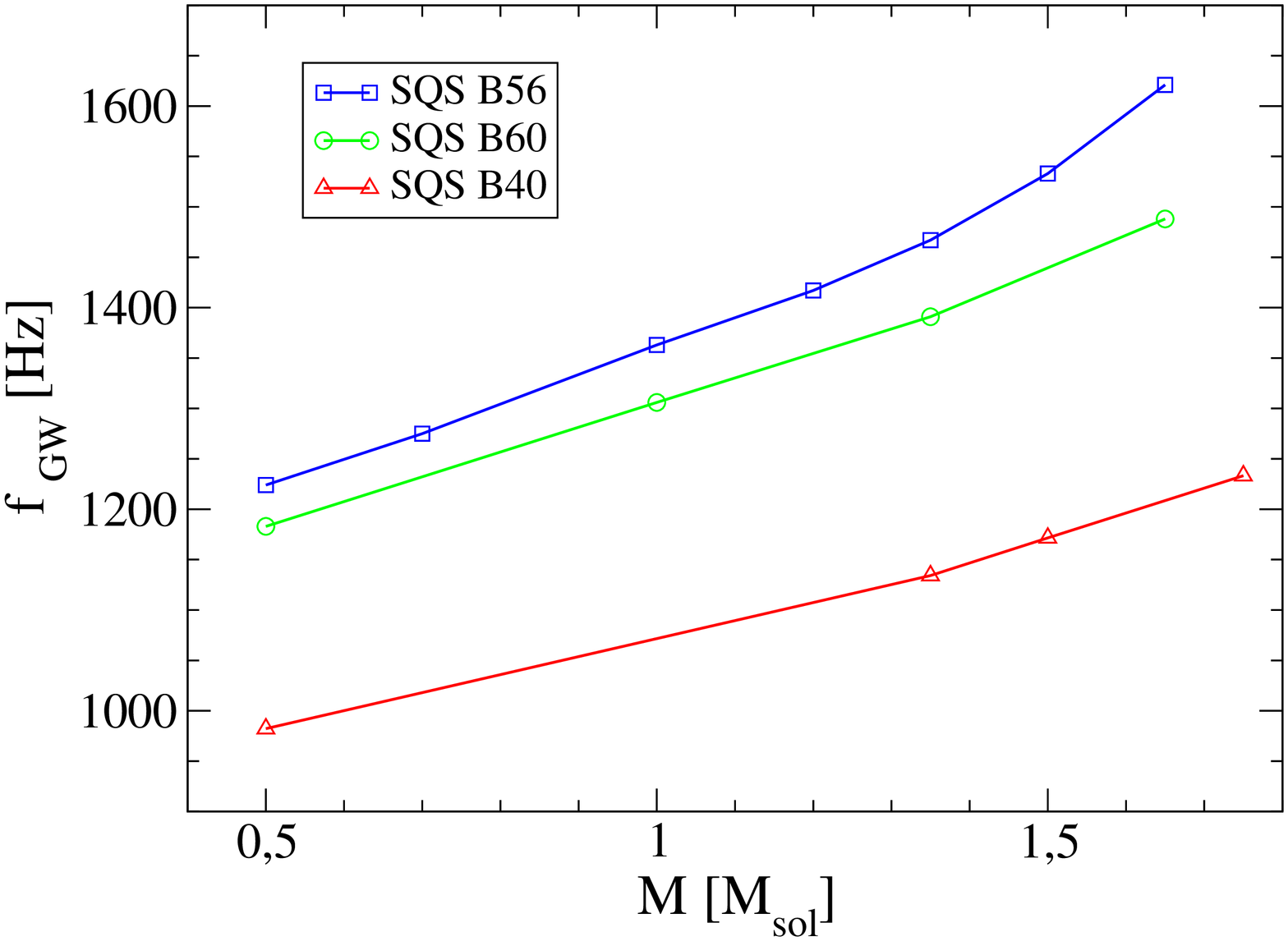}
\includegraphics[angle=-90,clip,width=0.5\textwidth]{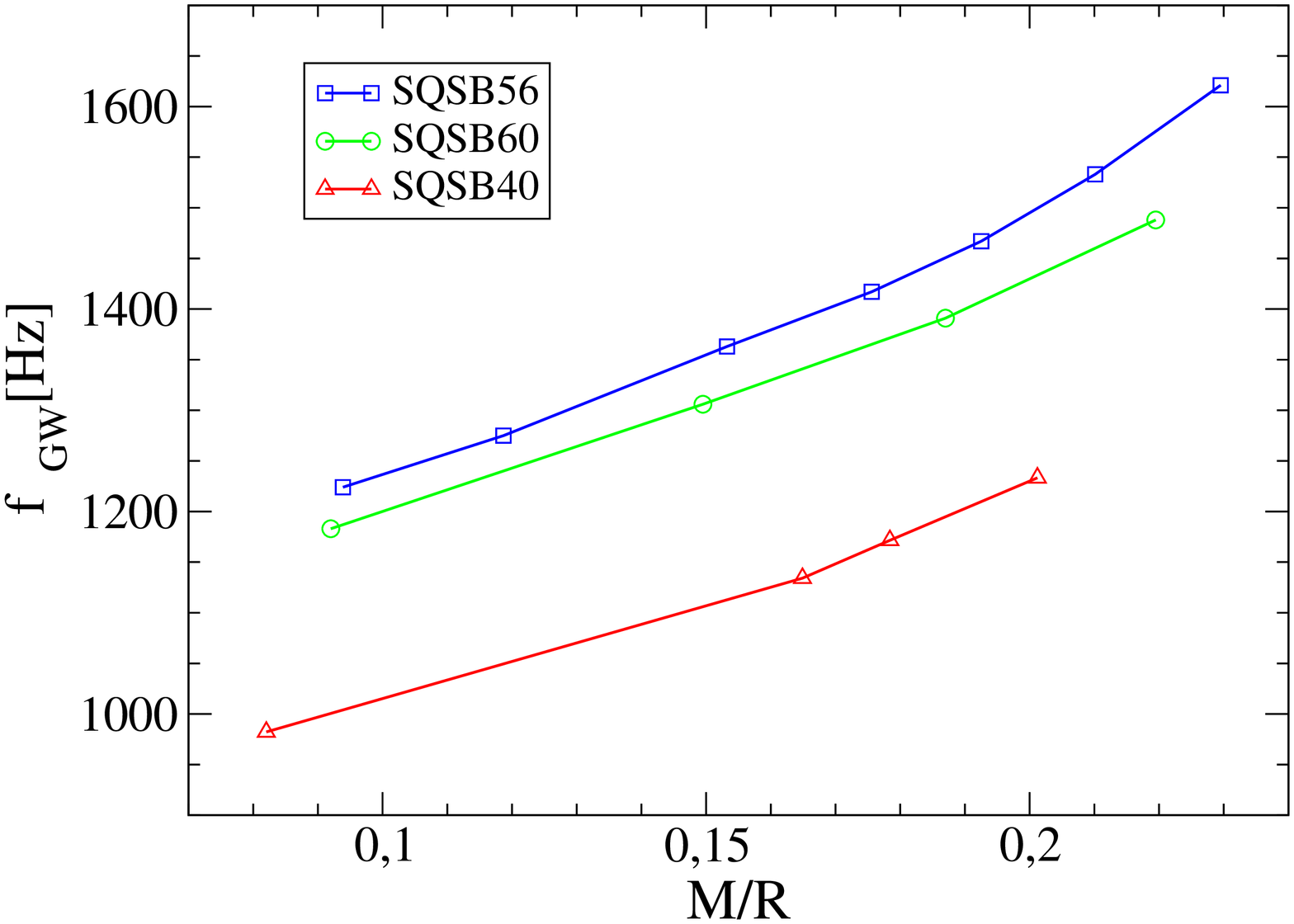}
\caption{The upper (lower) panel corresponds to the dependence of
  gravitational wave frequency at the ISCO on gravitational mass (compaction
  parameter) of a star in isolation for equal-mass binaries described by
  different strange quark matter EOS }
\label{f:fisco_M}
\end{figure}

\begin{table}[h]
\begin{center}
\begin{tabular}{|c|c|c|c|}
\hline
$M[M_{\odot}]$ &$A~[M_{\odot}]$ & $n$ & $f_{\rm npm}$\\
\hline\hline
1 & 0.0003781 & 6.02 & 593 \\
\hline
1.35  & $\ $0.0002888 $\ $& $\ $6.59$\ $ & 705 \\
\hline
1.65 & 0.0001590 & 7.27 & 838 \\
\hline
\end{tabular}
\end{center}
\caption{Parameters $A$ and $n$ of polynomial fits (\ref{e:fit}) for different
 masses evolutionary sequences of strange stars described by SQSB60 model. }
\label{fittable_diffmass}
\end{table}

In the upper (lower) panel of Fig. \ref{f:fisco_M} we show the frequency of GW
at the ISCO versus gravitational mass (the compaction parameter) for  three MIT bag
models SQSB56, SQSB60 and SQSB40. Different symbols (triangles, circles and
squares) correspond to the ISCO defined by the orbital instability. We find
that for all MIT bag models considered in the paper the frequency of GW at the
ISCO increases with the stellar mass and the compaction parameter
$M/R$ (the relation gravitational mass vs compaction parameter is
almost linear for the range of masses considered here).

\begin{figure}{}
\includegraphics[angle=-90,width=0.5\textwidth]{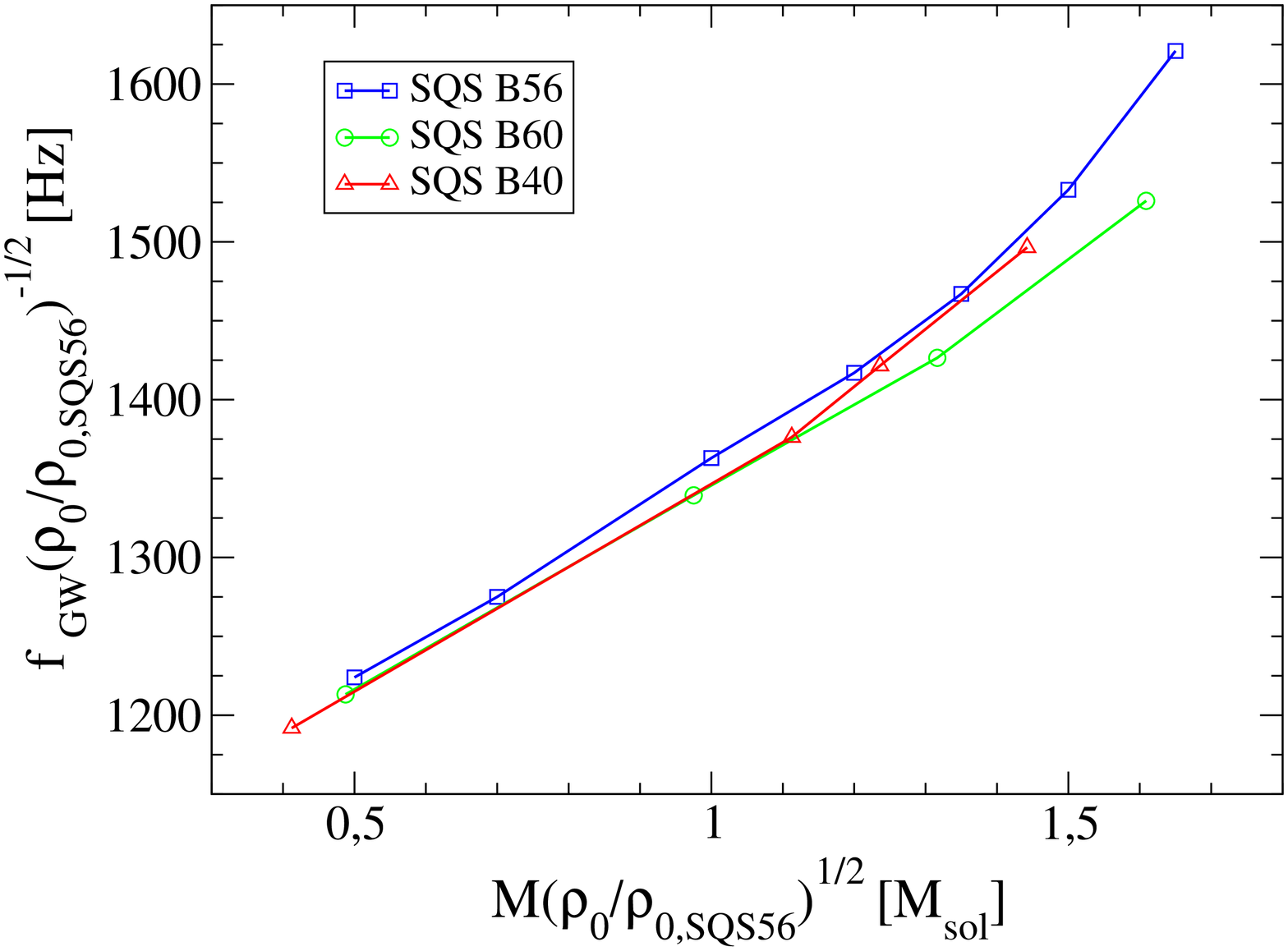}
\vskip 1cm
\includegraphics[angle=-90,width=0.5\textwidth]{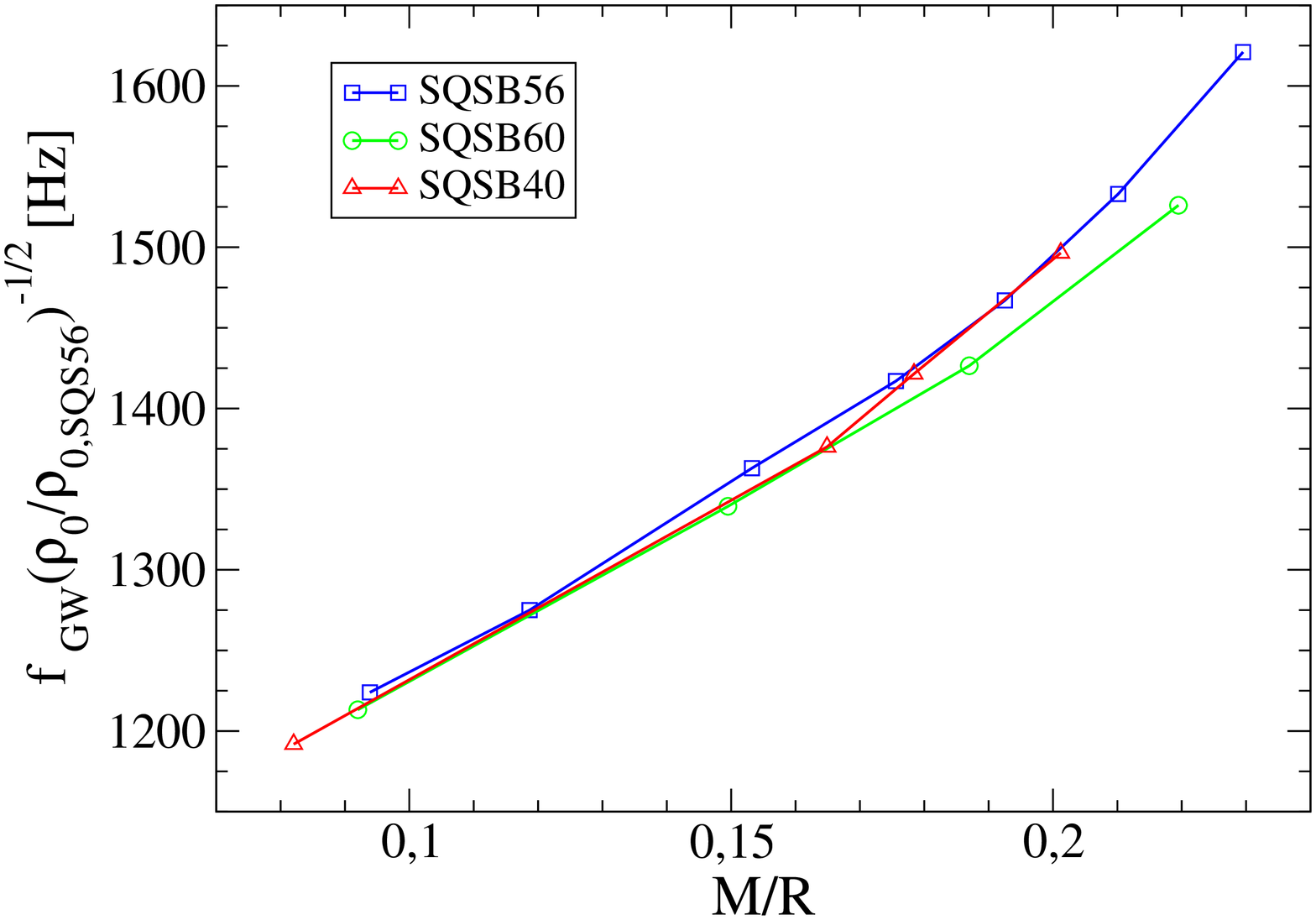}
\caption{Upper panel (lower panel) represents the rescaled gravitational wave 
frequency at the ISCO as a function of the rescaled gravitational mass 
(compaction parameter) of a star in isolation.}
\label{f:fisco_ren}
\end{figure}

\begin{figure}{}
\includegraphics[angle=-90,width=0.5\textwidth]{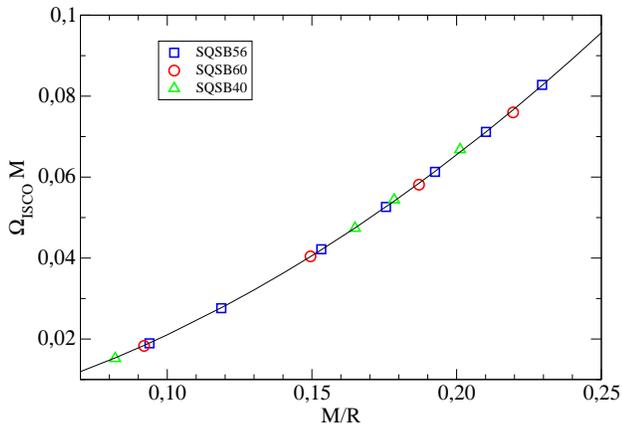}
\caption{The GW frequency at the ISCO multiplied by the gravitational
  mass of a strange star at isolation versus compaction parameter for
  equal-mass strange star binaries. The solid line correspond to the fitting
  formulae, while different symbols correspond to numerical results for different
  strange star models. }
\label{f:fMSS}
\end{figure}

\begin{figure}{}
\includegraphics[angle=-90,width=0.5\textwidth]{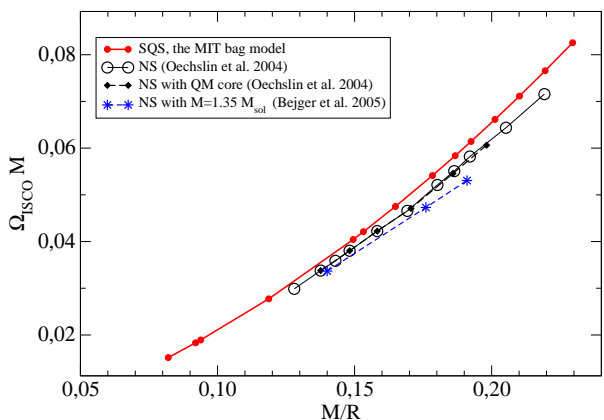}
\caption{The GW frequency at the ISCO multiplied by the gravitational
  mass versus compaction parameter for equal-mass neutron star and
  strange star binaries. Lines correspond to the fitting
  formulas, while different symbols correspond to numerical results for different
  strange star and neutron star models.}
\label{f:fMall}
\end{figure}

It was already mention that stars built predominantly of strange quark matter
described by the MIT bag model or the Dey et al. (1998) model can be very well
approximated by the linear function $P = a\left(\rho - \rho_0\right)$. For a
fixed value of $a$, all stellar parameters are subject to scaling relations
with appropriate powers of $\rho_0$, e.g. the gravitational mass and the
stellar radius scale in the same way: $M$, $R \propto \rho_0^{-1/2}$ while the
rotational frequency $f_{\rm rot} \propto \rho_0^{1/2}$ (e.g. \cite{Witten84,
Zduni00}). In Fig. \ref{f:fisco_ren} the dependence of the frequency of GW at
the ISCO versus mass and compaction parameter was scaled out with appropriate
powers of $\rho_{\rm 0,SQSB56}$, where $\rho_{\rm 0,SQSB56}=4.4997 \ [10^{14}{\rm
g/cm^3} ]$ corresponds to the surface density for the standard MIT bag
model SQSB56.  We see that the functions $f_{{\rm GW}}(M)$ and $f_{{\rm
GW}}(M/R)$ weakly depend on the $a$ parameter of the equation of state given
by Eq. (1). 

In Fig. \ref{f:fMSS} and \ref{f:fMall} we present the dimensionless quantities, the gravitational waves frequency at the ISCO multiplied by the gravitational mass of a  star in isolation  $M\Omega$  as a
function of $M/R$ for equal-mass  strange star binaries described by
the MIT bag model and  neutron star binaries
respectively.  The frequency at the ISCO depends systematically and
strongly on the compaction parameter for all models. This strong
dependence for binary neutron stars and strange stars indicates that
the effect of the hydrodynamic instability dominates over the general
relativistic effects (see \cite{UryuSE00} for a discussion on the
origin of ISCO) for compaction parameters considered in the paper. On
the other hand, in the limit $M/R \rightarrow 0.5$, the general
relativistic effects should be more important than the hydrodynamic
effects.

We find that the dependence of the frequency of gravitational waves at
the ISCO on the compaction parameter for the equal-mass binaries can
be described by the same simple analytical formulae (given by the
power series of the compaction parameter) $$\Omega M =A
(M/R)^{1.5}+B(M/R)^{2.5}$$ (see
e.g. \cite{UryuSE00,Baumgarte01}),where $A=0.6 $ and $B=0.665$, for
broad ranges of masses independently on the strange star model.  The
results obtained for one strange quark model can then be used to
predict the results for other models. Having this in mind we can
estimate that the gravitational wave frequency at the ISCO $f_{\rm
  GW,ISCO}$ for DSQS with $M_\infty=2.7$ could be very high $\sim 2.6$
kHz.

Comparison of our numerical results for three different models of
strange stars, represented by the solid line with filled circles, and
irrotational neutron star binaries are given in Fig \ref{f:fMall}.  The
solid line with open circles and the dashed line with filled squares
correspond to equal-mass neutron stars of Oechslin et al. 2004
\cite{Oechslin04} described by a pure nuclear matter EOS, based on a
relativistic mean field model and a `hybrid' EOS with a phase
transition to quark matter at high density respectively. These two
EOSs are matched with a polytropic one with high adiabatic index
$\gamma=2.86$ at $2\times 10^{14}\ {\rm g\, cm}^{-3}$.  This last
assumption of Oechslin et al. is artificial, because the EOS of the
neutron star crust is very different from a polytrope, and its local
adiabatic index is much smaller (see \cite{Bejger05} for details). The
dashed line with stars denotes results obtained by Bejger at al. 2005,
for binary neutron stars of equal masses with $M_\infty=2.7 M_\odot$
based on realistic equations of state for the whole neutron star
interior (see section V. A of this paper). They performed calculations
for three realistic nuclear EOSs of various softness at stellar core
and the crust described by means of a realistic EOS obtained in the
many-body calculations (see \cite{Bejger05} for details).  In this
case ISCO is defined by the point where mass transfer sets in.  In
contrast for irrotational strange star binaries and neutron star
binaries of Oechslin et al. the ISCO is given by the dynamical
instability due to very high adiabatic index in the stellar crust (see
\cite{LimouGG05}). The difference in the relation $\Omega M (M/R)$
between two models of Oechslin et al., caused by difference of EOS in
the stellar core, are negligible comparing to difference between their
results and our numerical calculations or results obtained by Bejger
et al. The last orbits of inspiral of binary neutron stars and strange
stars are dominated by tidal effects.  We conclude that the difference
between different models of compact stars, presented on
Fig. \ref{f:fMall} come from different description of the nuclear
crust.

\section{Summary}

In the present paper we have computed the final phase of inspiral of
equal-mass irrotational binary stars built predominantly of strange quark
matter. We have studied the precoalescing stage within a
quasiequilibrium approximation and a conformally flat spatial 3-metric
using a multidomain spectral method. We have presented a set of
evolutionary sequences of equal-mass strange star binaries based on
two types of equation of state at zero temperature, the MIT bag model
and the Dey et al. (1998) model, of strange quark matter. For each
sequence we have computed the gravitational waves energy spectrum.  We
have compared our results with those obtained for neutron star
binaries and the third order Post-Newtonian point-mass binaries.  We
have studied the impact of the equation of state and the total
energy-mass on the last orbits of binary strange quark stars by
finding the gravitational wave frequency at the ISCO, which marks the
end of the inspiral phase, and the break frequencies (GW frequencies
at which the energy spectrum drops by some factor below the point-mass
result) for each evolutionary sequence. These frequencies could be
determined from data analysis and allow us to make constraints on the
equation of state of neutron stars.

We find  that:

i) for equal-mass irrotational strange quark star binaries ISCO is
given by the dynamical instability independently on the equation of
state and the total energy-mass of the system. This contrasts with
neutron stars described by nuclear equation of state for which the
ISCO is given by mass-shedding limit.  The gravitational wave
frequency at the ISCO is always higher than 1.1 kHz for irrotational
strange quark stars described by MIT bag model and 2 kHz for the Dey
et al. (1998) model with the total mass-energy of a binary system
greater than $2 M_\odot$. One should note here that for non-equal mass
binaries a star of smaller mass could be tidally disrupted by a
companion of larger mass at large orbital separation and than the
frequency of gravitational waves could be smaller than 1 kHz.

ii) the frequency of gravitational waves at the ISCO depends
systematically and strongly on the compaction parameter (for $M/R <
0.24 $) for all models of strange quark stars. This indicates that the
ISCO is determined by the hydrodynamic instability and not by general
relativistic effects. The dependence of the frequency of gravitational
waves at the ISCO on the compaction parameter for the equal mass
binaries can be described by the same simple analytical formulae for
broad ranges of masses independently on a strange star model. The
higher the compactness of a star is, the higher the frequency of GW at
the ISCO is.

iii) the range of GW frequencies, $[1130, 1470]$, at the ISCO for
binary strange stars of $2.7 M\odot$ total mass, described by the MIT
bag model intersects with the range of frequencies, $[806, 1270]$, for
binary neutron stars. The determination of the gravitational wave
frequency at the ISCO by the laser interferometers wouldn't be
sufficient to distinguish without ambiguities between strange stars
and neutron stars.  It would be necessary to take into account the
observed deviation of the gravitational energy spectrum of a
quasiequilibrium sequence from point-mass behavior (the break
frequencies). The fits of the deviation between numerical results from
3PN results show that the power is very high, $n> 6$, which indicates
that high order tidal effects are very important.

iv) The frequency of GW
at the end point of inspiraling neutron stars described by several realistic
EOS can be predicted, in a good approximation, by studying binaries with
assumed polytropic EOSs with $\gamma = 2$ or $2.5$ with the same compaction
parameter. In contrast, the frequency of GW at the ISCO is always higher for
strange stars binaries than for polytropic neutron stars binaries with the
same compaction parameter. The differences in the evolution of binary strange
stars and neutron stars stem from the fact that strange stars are principally
bound by an additional force, strong interaction between quarks.

v) the higher total mass of the system is the higher frequency of GW
at ISCO. For MIT bag model the frequency of gravitational waves at the
ISCO only weakly depends on the parameter $a$ of the EOS, especially
for small compaction parameter.  The results obtained for one model
can thus be used for other MIT bag model using the scaling with
$\rho_0$.

In future work we plan to study binary neutron stars (strange quark
stars) with different mass ratio, e.g. 0.7, following the results of
\cite{BulikGB04} for the observability weighted distribution of double
neutron star binaries as well as binary systems consisting of one
strange star and one neutron star.

\acknowledgements 
We are grateful to Koji Uryu, Thomas Baumgarte and
Keisuke Taniguchi for helpful discussions and valuable comments.  This
work was supported, in part, by the KBN grant 1P03D00530,``Ayudas para
movilidad de Profesores de Universidad e Invesigadores espanoles y
extranjeros'' from the Spanish MEC, and by the Associated European
Laboratory Astro-PF (Astrophysics Poland-France).


\end{document}